\documentclass[aps,pra,twocolumn,groupedaddress,nofootinbib,notitlepage,
superscriptaddress,showpacs,floatfix]{revtex4-1}
\usepackage{graphicx,graphics,epsfig,subfigure,times,bm,bbm,amssymb,amsmath,
amsfonts,amsthm,mathrsfs,MnSymbol}
\usepackage[matrix,frame,arrow]{xypic}
\usepackage[pdfstartview=FitH]{hyperref}
\usepackage{pifont}
\hypersetup{
    colorlinks=true,       	
    linkcolor=red,          	
   citecolor=magenta,        
    filecolor=magenta,      	
    urlcolor=cyan,           	
    runcolor=cyan
}
\usepackage[pdftex]{color}
\usepackage{braket}
\usepackage{enumerate}
\usepackage[normalem]{ulem}

\usepackage{subfigure}
\usepackage{overpic}

\usepackage{multirow}
\newcommand{\be}{\begin{equation}}
\newcommand{\ee}{\end{equation}}
\newcommand{\ba}{\begin{eqnarray}}
\newcommand{\ea}{\end{eqnarray}}

\newcommand{\ignore}[1]{}

\begin{document}

\title{Quantum Computation in Noiseless Subsystems with Fast Non-Abelian Holonomies}

\author{J. Zhang}
\email{zhangjiangsd@foxmail.com}
\affiliation{Department of Physics, Shandong University, Jinan
250100, China}
\affiliation{Centre for Quantum Technologies, National University of
Singapore, 3 Science Drive 2, Singapore 117543}
\author{ L.-C. Kwek}
\email{cqtklc@nus.edu.sg}
\affiliation{Centre for Quantum Technologies, National University of
Singapore, 3 Science Drive 2, Singapore 117543}
\affiliation{National Institute of
Education, Nanyang Technological
University, 1 Nanyang Walk, Singapore 637616}
\affiliation{Institute of Advanced Studies, Nanyang
Technological University, 60 Nanyang View, Singapore 639673}
\author{Erik Sj\"{o}qvist}
\email{erik.sjoqvist@kemi.uu.se}
\affiliation{Centre for Quantum Technologies, National University of
Singapore, 3 Science Drive 2, Singapore 117543}
\affiliation{Department of Quantum Chemistry, Uppsala University, Box 518,
Se-751 20 Uppsala, Sweden}
\author{D. M. Tong}
\email{tdm@sdu.edu.cn}
\affiliation{Department of Physics, Shandong University, Jinan
250100, China}
\author{P. Zanardi}
\email{zanardi@usc.edu}
\affiliation{Centre for Quantum Technologies, National University of
Singapore, 3 Science Drive 2, Singapore 117543}
\affiliation{Department of Physics and Astronomy and Center for Quantum Information
Science \& Technology, University of Southern California, Los Angeles, CA 90089-0484}

\date{\today }

\begin{abstract}
Quantum information processing requires a high degree of isolation from the detrimental
effects of the environment as well as an extremely precise level of control on the way quantum
dynamics unfolds in the information-processing system. In this paper, we show how these two
goals can be ideally achieved by hybridizing the concepts of noiseless subsystems and of
holonomic quantum computation. An all-geometric universal computation scheme based
on non-adiabatic and non-Abelian quantum holonomies embedded in a four-qubit
noiseless subsystem for general collective decoherence is proposed. The implementation
details of this synergistic scheme along with the analysis  of its stability against
symmetry-breaking imperfections are presented.
\end{abstract}

\maketitle

\section{Introduction}
Implementation of quantum information processing (QIP) poses daunting challenges. In the
first place, for most of the QIP protocols, quantum coherence has to be maintained throughout
the whole computational process in spite of the decoherence induced by the unavoidable
coupling with  environmental degrees of freedom. Secondly, one has to achieve an unprecedented
level of control to enact   quantum gates within  the required high accuracy.

To the aim of accomplishing these, somewhat contradictory, tasks several theoretical schemes
have been devised since the early days of QIP. Broadly speaking, all the information-stabilizing
strategies developed to date fall in three categories: active techniques like quantum error correcting
codes \cite{qecc}, symmetry-aided passive ones like decoherence-free subspaces and subsystems
\cite{dfs,KLV,stab}, and geometrical \cite{zanardi99, jones00,duan01} and topological ones
\cite{pachos-topo}.

Geometric QIP exploits different types of quantum holonomies, e.g., Berry phases, to implement
quantum gates. Following the first non-Abelian \cite{zanardi99}  and  Abelian  \cite{jones00}
adiabatic proposals, many others have been considered, see, e.g., Refs.
\cite{duan01,wu05,cen06,zhang06,feng09}. The motivating idea is that the geometric nature
of the proposed quantum gates endows them with some degree of inherent robustness against
control imprecisions as well as against environment-induced decoherence \cite{hqc-robust}.
One of the  drawbacks of the original holonomic quantum computation (HQC) \cite{zanardi99}
is its  relative slowness due to the adiabaticity constraint. This potential limitation can be
circumvented by resorting to non-adiabatic Abelian \cite{wang01,zhu03} and non-Abelian
\cite{sjoqvist12,xu12} quantum holonomies.

The idea of noiseless subsystem (NS) was first introduced in Ref. \cite{KLV} and experimentally
demonstrated in Ref. \cite{NS-exp}. NSs are a natural generalization of the concept of noiseless
quantum code or decoherence-free subspace (DFS) \cite{dfs} and are effective when the
decohering interactions  possess  some non-trivial algebra of symmetries. On general
theoretical grounds, NSs have been argued to provide the unified algebraic structure underlying
all the known quantum-information protection schemes \cite{stab} including topological quantum
computation \cite{top}.

The goal of this paper is to merge synergistically  ideas from  geometric QIP and NSs  in order
to take advantage of the appealing features of both.  More specifically, we will hybridize
non-adiabatic  HQC \cite{sjoqvist12},  with the powerful theory of NSs \cite{KLV, stab}. The
possibility of achieving robust quantum control on NSs by non-Abelian quantum holonomies
was first envisioned in Refs. \cite{virt,oreshkov09},  universal HQC schemes embedded in DFSs
and NSs were  proposed in Refs. \cite{wu05,xu12} and (for a strongly dissipative case) in Ref.
\cite{angelo-DFS-HQC}.

In this paper, we extend significantly the results of Ref. \cite{xu12} by showing how a universal
non-Abelian and nonadiabatic holonomic processor can be embedded within a NS for general
collective decoherence. In this way a universal computational scheme protected against general
collective decoherence and featuring at the same time the robustness of HQC  against imprecisions
in gate control can be ideally achieved.

\section{Noiseless Subsystem}
We start by briefly recalling the basic notions concerning NSs. Let $\mathcal H$ be the Hilbert
space of a quantum system $S$ coupled to its environment through a set of ``error" operators
$\{E_\alpha\}_\alpha\subset {\mathcal B}({\mathcal H})$ \cite{bib_footnote}. The key object is
provided by the interaction algebra ${\mathcal A}\subset {\mathcal B}({\mathcal H})$, i.e., the
$C^*$-algebra \cite{C*}  generated by the error operators. The state space of the system decomposes
into the different $d_J$-dimensional irreducible representations (irreps) of $\mathcal A$ (labeled by $J$
and with multiplicity $n_J$)  as ${\mathcal H} \cong \oplus_J {\mathbb{C}}^{n_J} \otimes
{\mathbb{C}}^{d_J}$. The corresponding orthogonal decomposition of $\mathcal A$ is given by
\begin{equation}
{\mathcal A} \cong \oplus_J  {\mathbf{1}}_{n_J}\otimes M_{d_J} ,
\label{split}
\end{equation}
where $M_{d_J}$ denotes the full matrix algebra of $d_J\times d_J$ complex matrices \cite{C*}.
When $d_J=1$, one recovers the concept of DFS \cite{dfs}. The interaction algebra $\cal A$ acts
irreducibly on the ${\mathbb{C}}^{d_J}$ factors of $\mathcal H,$ whereas Eq. (\ref{split})  shows that
the error algebra elements, responsible for decoherence, have a {\em trivial} action on the ${\mathbb{C}}^{n_J}$ factors. It follows that quantum information can be protected by encoding
in these virtual subsystems \cite{virt}  that are then termed {\em noiseless subsystems} (NS) \cite{KLV}.
In order to perform manipulations of the NS-encoded information, one has to resort to a
non-trivial set of operations that belong to the commutant algebra ${\mathcal A}^\prime :=
\{ X\in {\mathcal B}({\mathcal H})\,/\, [X,\,A]=0, \forall A\in{\mathcal A}\}$. This crucial fact
follows from the dual irreps decomposition ${\mathcal A}^{\prime} \cong \oplus_J  M_{n_J}
\otimes {\mathbf{1}}_{d_J}$ for which one sees that of ${\mathcal A}^\prime$ has irreducible
action on the NSs and trivial one on the ${\mathbb{C}}^{d_J}$ factors, now playing the role of
multiplicity spaces. These constructions are useful if at least one of the $n_J$'s is larger than
or equal to two, which in turn gives a lower bound on the dimension of the commutant
${\mathcal A}^\prime$. In other words, the existence of a NS encoding relies on the existence
of a sufficiently large number of symmetries of the interaction algebra, i.e., of the noise.

The prototypical symmetric noise is provided by  collective decoherence \cite{dfs} whose
experimental relevance has been demonstrated in Refs. \cite{NS-exp, DFS-exp1, DFS-exp2}. This
is also the model that will be considered in this paper. In this collective case the interaction
algebra $\mathcal A$ is given by the algebra of totally symmetric operators on the state space
of $N$ qubits, i.e., ${\mathcal H}\cong ({\mathbb{C}}^2)^{\otimes\,N}$ and its commutant
${\mathcal A}^\prime$ is the $C^*$-algebra generated by permutations $\sigma\in{\cal S}_N$
acting on ${\mathcal H}$ according to the natural representation, i.e., $\sigma\colon \otimes_{p=1}^N |
\alpha_p \rangle\rightarrow \otimes_{p=1}^N |\alpha_{\sigma(p)} \rangle$. In the following, we
show how to enact a universal set of operators in ${\mathcal A}^\prime$ using non-adiabatic
quantum holonomies only.

\section{Non-adiabatic Holonomic Quantum Computation}
Non-adiabatic HQC, proposed in Ref. \cite{sjoqvist12} and experimentally implemented
in Ref. \cite{abdumalikov13}, is based on the concept of non-adiabatic non-Abelian
geometric phases \cite{anandan88}. The key idea is to implement a
suitably designed Hamiltonian that induces cyclic evolution of a quantum computational
system encoded in a subspace $\mathcal{M}(0)$ in such a way that all dynamical phases
vanish. The primitive structure is of $\Lambda$ type, where an excited state $\ket{e}$ is
coupled by a pair of simultaneous laser pulses to ground state levels $\ket{g_0},\ket{g_1}$
according to ($\hbar = 1$ from now on)
\begin{eqnarray}
H(t) = \Omega (t) \left( \omega_0 \ket{e} \bra{g_0} + \omega_1 \ket{e} \bra{g_1} +
\textrm{h.c.} \right) .
\label{eq:lambda}
\end{eqnarray}
Here, $\Omega (t)$ is the Rabi frequency and $\omega_0,\omega_1$ are complex-valued
time-independent driving frequencies satisfying $|\omega_0|^2 + |\omega_1|^2 = 1$. The
generic Hamiltonian $H(t)$ describes transitions between energy levels induced by oscillating
laser fields in the rotating wave approximation and can be implemented in a wide range of
different physical systems.

The subspace spanned by $\ket{\psi_j (t)} = e^{-i\int_0^t H(t') dt'} \ket{g_j} = U(t,0) \ket{g_j}$,
$j=0,1$, undergoes a cyclic evolution if the Rabi frequency satisfies $\int_0^{\tau} \Omega (t') dt'
= \pi$. The resulting time evolution operator $U(\tau,0)$, projected onto the computational
subspace $\mathcal{M}(0)=\textrm{Span} \{ \ket{g_0},\ket{g_1} \}$, defines the traceless Hermitian
gate $U(C_{\bf n}) = {\bf n} \cdot \boldsymbol{\sigma}$, where ${\bf n} = (\sin \theta \cos \phi,
\sin \theta \sin \phi , \cos \theta)$ with $\omega_0/\omega_1 = -\tan (\theta /2) e^{i\phi}$, and
$\boldsymbol{\sigma}$ are the standard Pauli operators acting on $\mathcal M(0)$. An arbitrary
SU(2) can be realized by sequentially applying two such gates with different ${\bf n}$. The evolution
is purely geometric since $\bra{\psi_j (t)} H(t) \ket{\psi_k (t)}$, $j,k=0,1$, vanish for $t\in [0,\tau]$.
Thus, $U(C_{\bf n})$ is fully determined by the path $C_{\bf n}$ of $\mathcal M(t)$ in the space of
all two-dimensional subspaces of the three-dimensional Hilbert space, i.e., in the complex-valued
Grassmannian $G(3;2)$. Together with an entangling holonomic two-qubit gate,
$U(C_{\bf n})$ constitutes a universal all-geometric set of quantum gates \cite{bremner02}.

\section{Quantum holonomy in noiseless subsystems}
The collective decoherence on a quantum system $S$ consisting of $N$ physical qubits is
characterized by the spin-$\frac{1}{2}$ error operators $E_{\alpha}=\sum_{p=1}^{N}\sigma_p^{\alpha}$,
$\alpha=\pm,z$. For a fixed total spin $J$, the dimension of the noiseful (NF) part is $d_{J}=2J+1$.
By using angular momentum addition rules, one can prove that
\begin{equation}
n_{J}=\frac{(2J+1)N!}{(N/2+1+J)!(N/2-J)!} .
\end{equation}
This $n_{J}$, which is the dimension of the NS part, provides the possibility of performing HQC.

Quantum holonomy appears when the subspace $\mathcal{M} (t)$ returns to the original
one after a non-trivial cyclic transformation. The NS spans the total space, which is the
total Hilbert space of a HQC. In general, the NS should be larger than the computational
space in order to admit non-trivial holonomies. A subspace of NS emerges as $\mathcal{M} (0)$
and the effective Hamiltonian of NS acts as the Hamiltonian that generates the non-trivial loop
relating to the unitary transformation. Since NS theory guarantees that the states in $\mathcal{M} (0)$
are never evolved out of the NS, the subspace of NS can return to the original one and that assures
that HQC can be conducted.

\subsection{One-qubit gate} 
Non-adiabatic one-qubit holonomic gate can be implemented in
a NS provided there exists a $J$ for which $n_J \geq 3$. Four physical qubits, which contains a
$\mathbb{C}^{3} \otimes \mathbb{C}^{3}$ subspace, provides the smallest possible realization
of such a gate. Here, we demonstrate how noiseless holonomic one-qubit gates can be
implemented in this $\mathbb{C}^{3} \otimes \mathbb{C}^{3}$ subspace of the four-qubit code.

First, note that the $E_{\alpha}$'s act on the $\mathbb{C}^{3} \otimes \mathbb{C}^{3}$ subspace
in the following way,
\begin{eqnarray}
E_{\alpha} = I_{\textrm{NS}} \otimes S_{\alpha}, \ \alpha = \pm,z,
\end{eqnarray}
where $S_{\alpha}$ are spin$-1$ representation of the angular momentum operators.
An important observation here is that the inherent symmetry in the action of the decoherence
operators $E_{\alpha}$ on the basis states $\ket{i}\ket{j}_{4}$'s affects
only the second part of the basis and leaves the first part unchanged (see the Appendix 
for more details on the four-qubit code). Moreover, the basis changed
by the error operators stays within $\ket{i}\ket{j}_4$'s. Thus, the information being stored
in this subspace depends only on the first index -- it is therefore not spoilt by the interaction
between the system and the environment.

To perform holonomic one-qubit gates with the four-qubit code, a set of operators is
needed to achieve the appropriate transitions so that the computation stays within the subspace.
To this end, the operators that we seek should commute with the $E_{\alpha}$'s. Let us consider
the permutation operator $P_{pq}=\frac{1}{2}(I_{pq} + \boldsymbol{\sigma}_p \cdot
\boldsymbol{\sigma}_q)$ of qubits $p$ and $q$ such that $P_{pq}\ket{x}_{p}\ket{y}_{q} =
\ket{y}_{p}\ket{x}_{q}$ for $x,y\in\{0,1\}$. Here, $I_{pq}$ is the identity and $\boldsymbol{\sigma}_p,
\boldsymbol{\sigma}_q$ are the Pauli operators acting on this qubit pair. It is straightforward to
check that $[P_{pq},E_{\alpha}]=0$. Three- and four-body permutation operators emerge as a
product of two-body ones. Thus, if the Hamiltonian is constructed using a combination of the
permutation operators, it will not destroy the subspace. Explicitly, we may take the Hamiltonian
that generates the holonomic one-qubit gates to be
\begin{widetext}
\begin{eqnarray}
H^{(1)}(t) & = & \Omega (t) \left[ \frac{J_1}{\sqrt{3}}(P_{23}-P_{13}) + i \frac{J_2}{\sqrt{3}}(P_{23} P_{13} -
P_{13}P_{23}) + \frac{J_4}{2\sqrt{6}}(P_{13}-P_{23}-3P_{14}+3P_{24}) \right]
\nonumber \\
 & = & \Omega (t) \left[ \left( J_1 - i J_2 \right) \ket{3} \bra{1} + J_4 \ket{3} \bra{2} +
\textrm{h.c.} \right] \otimes I_{\textrm{NF}} ,
\end{eqnarray}
\end{widetext}
where the first tensor factor corresponding to the NS is identical to the Hamiltonian in Eq.
(\ref{eq:lambda}) by identifying $\ket{1} = \ket{g_0}$, $\ket{2} = \ket{g_1}$, and $\ket{3} =
\ket{e}$; $I_{\textrm{NF}}$ is the identity operator acting on the noiseful subsystem. The
Hamiltonian vanishes on the noiseless qubit subspace $\mathcal{M}^{(1)} (0) = \textrm{Span} \{ \ket{1},
\ket{2} \}$, which guarantees the geometric nature
of the evolution. By putting $(J_1-iJ_2)/J_4 = -\tan (\theta /2) e^{i\phi}$ and defining the unit
vector ${\bf n} = (\sin \theta \cos \phi ,\sin \theta \sin \phi ,\cos \theta )$, a traceless one-qubit
holonomic gate
\begin{eqnarray}
U^{(1)} (C) = {\bf n} \cdot \boldsymbol{\sigma} \otimes I_{\textrm{NS}}
\end{eqnarray}
acting non-trivially on the two-dimensional subspace of the NS can thus be implemented by choosing
$\int_0^{\tau} \Omega (t) dt = \pi$. By combining two such gates, an arbitrary SU(2) operation acting
on the noiseless qubit subspace $\mathcal{M}^{(1)} (0)$ can be realized.

\subsection{Two-qubit gate} It is well-known that universal quantum computation can be achieved
as long as all one-qubit gates and a single non-trivial two-qubit (entangling) gate is possible
\cite{bremner02}. Since all single qubit gates are possible, it remains to demonstrate that we
could construct a non-trivial two-qubit gate.

To guarantee the holonomic scheme to be scalable, we encode each qubit in a two-dimensional
subspace of a three-level NS by using four physical qubits. In this scheme, a two-qubit gate
requires an eight-qubit code where two noiseless qubits are represented by two sets $L,L'$ of
four physical qubits. By choosing an appropriate Hamiltonian for the eight physical qubits, we
demonstrate a holonomic CNOT gate that can entangle these noiseless qubits.

Consider the eight-qubit Hamiltonian expressed in terms of permutation operators as
\begin{eqnarray}
H^{(2)} (t) &= &\frac{\Omega(t)}{12}(P_{13}-P_{23}-3P_{14}+3P_{24})_{L}
\nonumber \\
 & &  \mbox{\hspace{-1cm}} \otimes[P_{23}-P_{13}-\frac{1}{2\sqrt{2}}
(P_{13}-P_{23}-3P_{14}+3P_{24})]_{L'} .
\end{eqnarray}
By re-expressing the two factors in terms of the NS+NF basis for each four-qubit set, we obtain
\begin{eqnarray}
H^{(2)} (t) &= & \Omega (t) \left( H_0 + H_1\right) \otimes I_{\textrm{NF}} ,
\end{eqnarray}
where
\begin{eqnarray}
H_0 & = & \frac{1}{\sqrt{2}} \ket{33} \bra{21} - \frac{1}{\sqrt{2}} \ket{33} \bra{22} + \textrm{h.c.},
\nonumber \\
H_1 & = & \frac{1}{\sqrt{2}} \ket{31} \bra{23} - \frac{1}{\sqrt{2}} \ket{32} \bra{23} + \textrm{h.c.},
\end{eqnarray}
and $I_{\textrm{NF}}$ is now the identity on the nine-dimensional noiseful subsystem of the
eight qubits. The two time-independent operators $H_0$ and $H_1$ vanish on the computational
two-qubit subspace $\mathcal{M}^{(2)}(0)=\textrm{Span} \{ \ket{11},\ket{12},\ket{21},\ket{22} \}$
of the NS, which assures the geometric nature of the evolution. Furthermore, $H_0$ and $H_1$
commute, which implies that
\begin{eqnarray}
e^{-i\int_0^{\tau} H^{(2)} (t') dt'} = e^{-i \pi H_0}  e^{-i \pi H_1} \otimes I_{\textrm{NF}},
\end{eqnarray}
by choosing $\int_0^{\tau} \Omega (t') dt' = \pi$. The second factor $e^{-i \pi H_1}$ acts
trivially on $\mathcal{M}^{(2)}(0)$ and can therefore be ignored. The holonomic gate $U^{(2)}(C)$
is the projection of the first factor $e^{-i \pi H_0}$ onto $\mathcal{M}^{(2)}(0)$ and reads
\begin{eqnarray}
U^{(2)}(C) & = & \left( \ket{11} \bra{11} + \ket{12} \bra{12} \right.
\nonumber \\
 & & \left. + \ket{21} \bra{22} + \ket{22} \bra{21}
\right) \otimes I_{\textrm{NF}} .
\end{eqnarray}
We see that $U^{(2)} (C)$ is a CNOT gate acting on $\mathcal{M}^{(2)}(0)$ which completes the
universal set of non-adiabatic holonomic gates in NSs.

\section{Robustness of gates}   
Our NS encoding allows for perfect protection in the ideal
collective decoherence case where the system-bath interactions are fully invariant under
arbitrary  permutations of the physical qubits. However, in realistic situations symmetry-breaking
interactions will be unavoidably present and spoil the ideal behavior. In order to investigate
the robustness of our scheme against such unwanted interactions we introduce a simple
decoherence  model with a single parameter $g$ that controls the degree of symmetry
breaking. The noise Lindblad operators are given by: $E_{\alpha}^{'} =
\sum_{p=1}^{4}e^{-p g}\sigma_{\alpha}^{p}$; clearly when $g=0$ one recovers the
permutational invariant collective decoherence.
Within the usual Born-Markov approximation the system evolution is dictated by the Lindblad
master equation
\begin{eqnarray}
\dot{\rho}(t)&=&-i[H_{s}(t),\rho]+\Gamma(E_{z}^{'}\rho E_{z}^{'} -
\frac{1}{2}\{E_{z}^{'}E_{z}^{'},\rho\})
\nonumber \\
 & &+\gamma(\bar{n}+1)(E_{-}^{'}\rho E_{+}^{'}-\frac{1}{2}\{E_{+}^{'}E_{-}^{'},\rho\})
\nonumber \\
 & &+\gamma\bar{n}(E_{+}^{'}\rho E_{-}^{'}-\frac{1}{2}\{E_{-}^{'}E_{+}^{'},\rho\}),
\label{ME}
\end{eqnarray}
where $\bar{n}$ is the temperature-dependent average number of quanta in the environment
($\bar{n} = 0$ at zero temperature), $\Gamma$ and $\gamma$ are the dephasing rate and
dissipation rate, respectively. Here, $H_{s}(t)$ is the qubit system Hamiltonian, which is generated
by linear combination of the permutation operators as described above. A square pulse with
magnitude $\Omega$ and duration $\pi / \Omega$ is used. 

As a  figure of merit to
quantify the  robustness of our logical gates, we adopt the gate fidelity $F$ defined as the
Bures-Uhlmann fidelity $F(\rho_{id},\rho_f) := {\mathrm{Tr}} \sqrt{\rho_f^{1/2} \rho_{id}
\rho_f^{1/2}}$ averaged over initial conditions. Here,  $\rho_{id}$ is the NS state obtained by
the ideal holonomic  gate, i.e., the actual one in presence
of collective decoherence only, and $\rho_f$ is the corresponding faulty one obtained by solving
Eq. (\ref{ME}) and tracing over the noiseful degrees of freedom. 

\begin{figure}
\includegraphics[width=105mm]{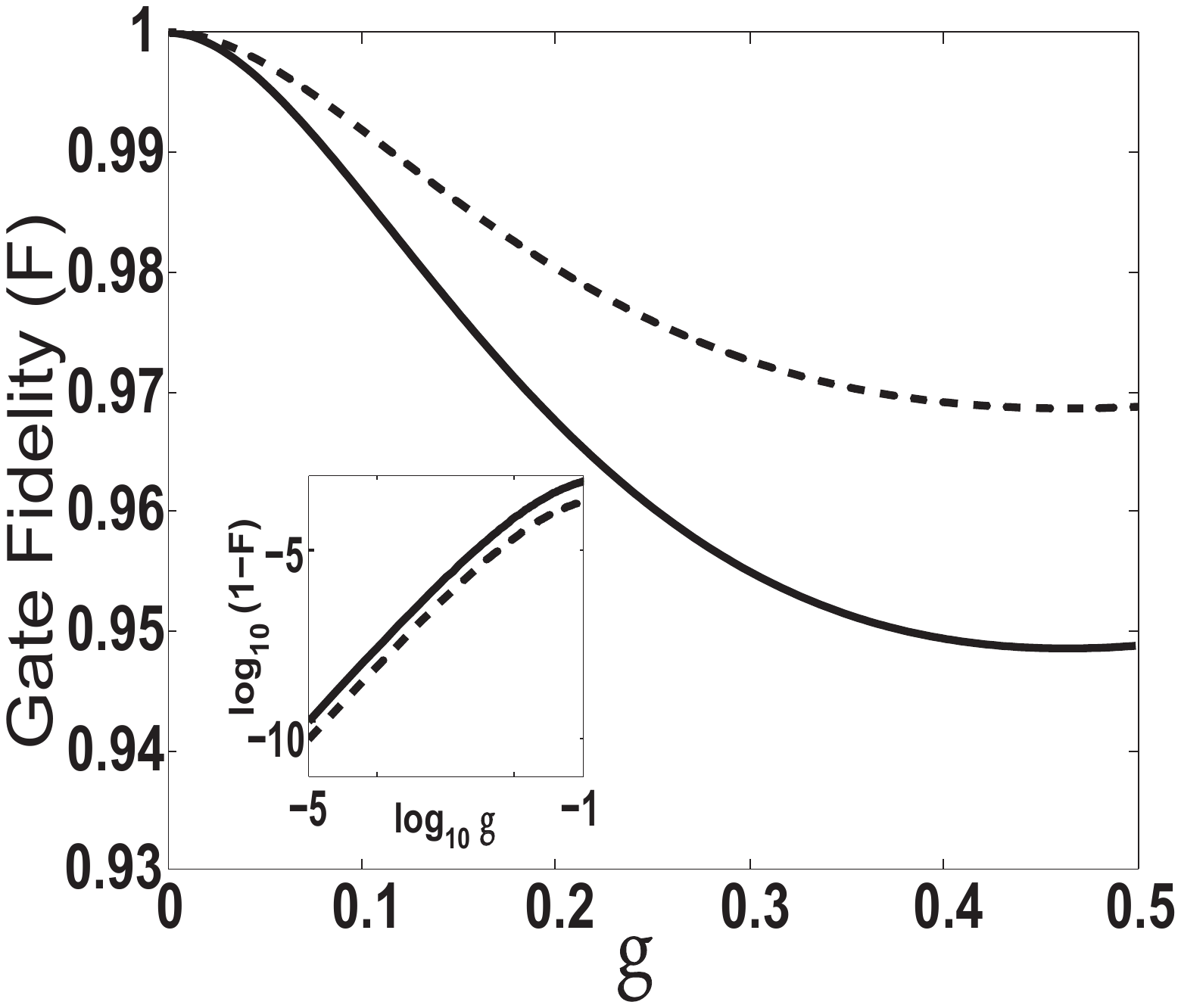}
\caption{\label{fg:1} Gate fidelity in presence of non-collective environment with $g$
controlling the degree of symmetry breaking. A square pulse with magnitude $\Omega$
and duration $\pi / \Omega$ is used. The dephasing rate $\Gamma$ and dissipation rate
$\gamma$ are chosen to satisfy $\Gamma = \gamma = 0.1 \Omega$. The logical
unitary gate is the standard Pauli Z operator and the fidelity is averaged over the six axial
pure states on the Bloch sphere as input states. The dashed line and the solid line show the
gate fidelity in NS for mean number of environmental quanta $\bar{n}=0$ and $\bar{n}=1$,
respectively. The inset shows the plots of $\log_{10} (1-F)$ as a function of $\log_{10} g$, 
dashed line and solid line are for $\bar{n}=0$ and $\bar{n}=1$, respectively.}
\end{figure}

We solved numerically Eq. (\ref{ME})
and  examined the gate fidelity as a function of $g$ for a one-qubit holonomic gate. In Fig.
\ref{fg:1} it is shown that gate fidelity, both at zero temperature (solid line; $\bar{n} = 0$) and
non-zero temperature (dashed line; $\bar{n} = 1$), decreases with increasing $g$ as expected.
However, the inset in Fig. \ref{fg:1} also shows that in the physically relevant regime of small $g$,
the gate fidelity behaves as $F\approx 1-g^{a}$, where $a\approx 2$. This demonstrates that
holonomic manipulations of the NS have some degree of resilience against slight violations of
the collective symmetry assumption. We would like to stress that Eq. (12)
 is just  a way to describe the system-bath coupling that allows one to interpolate, in a simple
phenomenological fashion, between the fully permutational symmetric ($g=0)$  and increasingly
non-symmetric and unprotected regimes (large $g$).  However,  we expect  the  conclusions
drawn from our simulations  to be generic for mild violations of permutational symmetry. Namely,
gate robustness should be independent of the details of the specific decoherence model, e.g.,
Markovianity.

\section{Conclusions}
In this paper, we have shown how to implement a universal set of one- and two-qubit gates
by non-adiabatic and non-Abelian quantum holonomies acting entirely within a noiseless
subsystem for general collective decoherence.  Each noiseless qubit can be encoded using
four physical qubits and geometrically manipulated by Heisenberg-like two- and four-body
interactions. The requested ability to enact four-body-interactions certainly  presents a major
challenge to the realization of our scheme with current experimental techniques. In order to
overcome this limitation one may think of resorting to geometric techniques to simulate many-body
interactions in terms of simpler interactions \cite{Wang-Simul} or to the  so-called perturbation
gadgets \cite{gadget}. In both cases ancillary degrees of freedom are needed. Finally,  by numerical
simulations, we have provided evidence of the robustness of the proposed hybrid scheme against
symmetry-breaking interactions with the environment.

\vspace{0.2cm}
\section*{Acknowledgments}
The work was financially supported by the National Research Foundation \& Ministry of
Education, Singapore. P.Z. is  supported by the ARO MURI grant W911NF-11-1-0268 and
by NSF grant  PHY- 969969. J. Z. and D. M. T. acknowledge support from NSF China with
No.11175105 and the Taishan Scholarship Project of Shandong Province. E.S. acknowledges 
financial support from the Swedish Research Council. 

\section*{Appendix}

The Hilbert space of a four-qubit system can be decomposed as
\begin{equation}
(\mathbb{C}^{2})^{\otimes4} = \mathbb{C}^2 \otimes \mathbb{C} \bigoplus \mathbb{C}^3 \otimes \mathbb{C}^3 \bigoplus \mathbb{C} \otimes \mathbb{C}^5 .
\nonumber
\end{equation}
By using the notation
\begin{eqnarray}
\ket{0} = \ket{1/2,1/2}, \ \ket{1} = \ket{1/2,-1/2},
\nonumber
\end{eqnarray}
we find the $\mathbb{C}^3 \otimes \mathbb{C}^3$ basis states
\begin{eqnarray*}
 & & \left\{\begin{array}{lll}
\ket{1}\ket{1}_{4} = \ket{1,1} & = & \frac{1}{\sqrt{6}}(2\ket{0010}-\ket{0100}-\ket{1000}), \\ 
\ket{1}\ket{2}_{4} = \ket{1,0} & = & \frac{1}{2\sqrt{3}}(2\ket{0011}-\ket{0101}-\ket{1001} \\
 & & + \ket{0110}+\ket{1010}-2\ket{1100}), \\
\ket{1}\ket{3}_{4} = \ket{1,-1} & = & \frac{1}{\sqrt{6}}(\ket{0111}+\ket{1011}-2\ket{1101}),
\end{array} \right. \\
\\
 & & \left\{ \begin{array}{lll}
\ket{2}\ket{1}_{4} = \ket{1,1} & = & \frac{1}{2\sqrt{3}}(3\ket{0001}-\ket{0010}-\ket{0100} \\
 & & -\ket{1000}), \\
\ket{2}\ket{2}_{4} = \ket{1,0} & = & \frac{1}{\sqrt{6}}(\ket{0011}+\ket{0101}+\ket{1001} \\
 & & - \ket{0110} - \ket{1010} - \ket{1100}), \\
\ket{2}\ket{3}_{4}=\ket{1,-1} & = & \frac{1}{2\sqrt{3}}(\ket{0111}+\ket{1011}+\ket{1101} \\
 & & -3\ket{1110}).
\end{array}\right. \\
\\
& &\left\{
\begin{array}{lll}
\ket{3}\ket{1}_4 = \ket{1,1} & = & \frac{1}{\sqrt{2}}(\ket{0100}-\ket{1000}), \\
\ket{3}\ket{2}_4 =\ket{1,0} & = & \frac{1}{2}(\ket{0101}-\ket{1001}+\ket{0110} \\
 & & -\ket{1010}), \\
\ket{3}\ket{3}_4 = \ket{1,-1} & = & \frac{1}{\sqrt{2}}(\ket{0111}-\ket{1011}),
\end{array} \right.
\nonumber
\end{eqnarray*}
The NS holonomies are realized in the first tensor factor of these $\ket{i}\ket{j}_{4}$ states.

The Gell-Mann matrices $\lambda_1,\ldots,\lambda_8$ on the NS in the $\mathbb{C}^3
\otimes \mathbb{C}^3$ subspace can be expressed in terms of qubit permutation operators as
\begin{eqnarray*}
\lambda_{1}\otimes I_{\textrm{NF}} & = &
\left( \ket{3} \bra{1} + \textrm{h.c.} \right) \otimes I_{\textrm{NF}}
                                 =\frac{1}{\sqrt{3}}(P_{23}-P_{13}), \\
\lambda_{2}\otimes I_{\textrm{NF}} & = &
\left( -i \ket{3} \bra{1} + \textrm{h.c.} \right) \otimes I_{\textrm{NF}},
                                 =i\frac{1}{\sqrt{3}}(P_{23}P_{13}-P_{13}P_{23}), \\
\lambda_{3}\otimes I_{\textrm{NF}} & = &
\left( \ket{3} \bra{3} - \ket{1} \bra{1} \right) \otimes I_{\textrm{NF}}
                                 =\frac{1}{3}(P_{13}+P_{23}-2P_{12}), \\
\lambda_{4}\otimes I_{\textrm{NF}} & = &
\left( \ket{3} \bra{2} + \textrm{h.c.} \right) \otimes I_{\textrm{NF}}
                                 =\frac{1}{2\sqrt{6}}(P_{13}-P_{23}-3P_{14}+3P_{24}), \\
\lambda_{5}\otimes I_{\textrm{NF}} & = &
\left( -i \ket{3} \bra{2} + \textrm{h.c.} \right) \otimes I_{\textrm{NF}}
                                =  i\frac{1}{2\sqrt{6}} (2P_{321}-2P_{231}+P_{342} \\
 & & -P_{432}-P_{341}+P_{431}+4P_{241}-4P_{421}), \\
\lambda_{6}\otimes I_{\textrm{NF}} & = &
\left( \ket{1} \bra{2} + \textrm{h.c.} \right) \otimes I_{\textrm{NF}}
                               = -\frac{1}{6\sqrt{2}}(2P_{13}+2P_{23}-4P_{12} \\
 & & +3P_{2341}+3P_{3421}+3P_{4321}+3P_{2431}-6P_{3241}-6P_{4231}), \\
\lambda_{7}\otimes I_{\textrm{NF}} & = &
\left( -i \ket{1} \bra{2} + \textrm{h.c.} \right) \otimes I_{\textrm{NF}} \\
                               & = & i\frac{1}{2\sqrt{2}}(P_{341}+P_{342}-P_{432}-P_{431}), \\
\lambda_{8}\otimes I_{\textrm{NF}} & = & \frac{1}{\sqrt{3}}
\left( \ket{3} \bra{3} + \ket{1} \bra{1} -2 \ket{2} \bra{2} \right) \otimes I_{\textrm{NF}} \\
               & = & \frac{1}{\sqrt{3}}(I-P_{12}-P_{13}-P_{23}).
\end{eqnarray*}
Note that the realization of  some of the Gell-Mann matrices ($\lambda_{2},\lambda_{5},\lambda_{6},\lambda_{7}$) requires higher than two-body interaction.


\begin{thebibliography} {99}
\bibitem{bib_footnote} The error operators may show up in the system-environment
interaction Hamiltonian, as Lindblad operators in master equation type of description,
or as Kraus operators in the resolution of the completely positive map describing the
finite-time decoherence process.
\end{thebibliography}
\end{document}